\documentclass{article}

\usepackage[preprint,nonatbib]{neurips_2022}

\usepackage[utf8]{inputenc} 
\usepackage[T1]{fontenc}    
\usepackage{hyperref}       
\usepackage{url}            
\usepackage{booktabs}       
\usepackage{multirow}
\usepackage{amsfonts}       
\usepackage{nicefrac}       
\usepackage{microtype}      
\usepackage{xcolor}         
\usepackage{graphicx}
\usepackage[normalem]{ulem}
\useunder{\uline}{\ul}{}

\title{BEANS: The Benchmark of Animal Sounds}

\author{%
  Masato Hagiwara, \ Benjamin Hoffman, \ Jen-Yu Liu, \ Maddie Cusimano \\ {\bf Felix Effenberger, \ Katie Zacarian} \\
  Earth Species Project \\
  \texttt{\{masato, benjamin, jenyu, maddie, felix, katie\}@earthspecies.org}
}

\begin{document}

\maketitle

\begin{abstract}
  The use of machine learning (ML) based techniques has become increasingly popular in the field of bioacoustics over the last years.
  Fundamental requirements for the successful application of ML based techniques are curated, agreed upon, high-quality datasets and benchmark tasks to be learned on a given dataset.
  However, the field of bioacoustics so far lacks such public benchmarks which cover multiple tasks and species to measure the performance of ML techniques in a controlled and standardized way and that allows for benchmarking newly proposed techniques to existing ones. 
  Here, we propose BEANS (the BEnchmark of ANimal Sounds), a collection of bioacoustics tasks and public datasets, specifically designed to measure the performance of machine learning algorithms in the field of bioacoustics.
  The benchmark proposed here consists of two common tasks in bioacoustics: classification and detection. 
  It includes 12 datasets covering various species, including birds, land and marine mammals, anurans, and insects. 
  In addition to the datasets, we also present the performance of a set of standard ML methods as the baseline for task performance.
  The benchmark and baseline code is made publicly available~\footnote{\url{https://github.com/earthspecies/beans}} in the hope of establishing a new standard dataset for ML-based bioacoustic research.
\end{abstract}

\section{Introductions}

\begin{figure}[h]
\begin{center}
\includegraphics[scale=0.35]{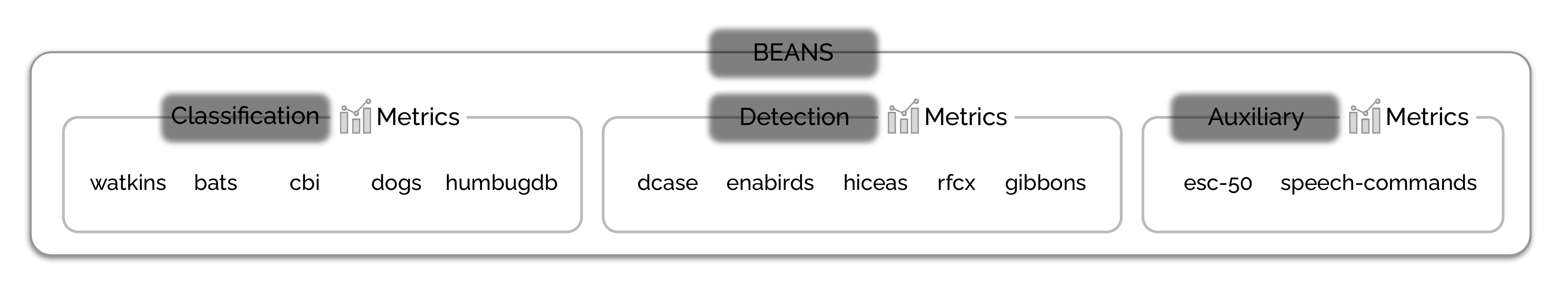} 
\caption{Overview of BEANS}
\label{fig:tasks}
\end{center}
\end{figure}

Due to their increasing affordability, recording and storage devices are now widely used to collect bioacoustic data. These devices enable animal welfare and wildlife conservation applications, such as passive acoustic monitoring (PAM), which offers tools for wildlife population assessment and conservation research in a non-invasive and unbiased manner~\cite{sugai2018pam}. However, they produce a large amount of bioacoustic data, and the manual processing and analysis of these data has become a bottleneck~\cite{stowell2022computational,tuia2022perspectives}.

For this reason, machine learning (ML) has increasingly been used to automate the processing and analysis of bioacoustics tasks. It has been successfully applied to a variety of tasks such as classification of species, individuals, and various other characteristics of calls~\cite{morfi2021deep}, detection and recognition of vocalizations in passive recordings~\cite{lebien2020pipeline}, and automatic discovery of vocalization units~\cite{sainburg2020latent}. Furthermore, the so-called ``deep learning'' (DL) models based on deep artificial neural networks have drastically reduced error rates and are increasingly being used for these tasks in recent years.

However, typical research studies in bioacoustics focus only on a small number of species and/or specific types of methodology~\cite{coffey2019deepsqueak,shiu2020deep,kahl2021birdnet}. This narrow focus has led to a proliferation of ML models and algorithms that perform well on the tasks in question on a given data set, but not necessarily outside of their scope~\cite{shwartz-ziv2022tabular}. Moreover, the lack of publicly available, agreed-upon standard datasets in bioacoustics~\cite{baker-vincent2019deafening,mclouphlin2019automated} makes it difficult to reproduce and compare different approaches in a standardized way.

If we turn our attention to other fields of machine learning, much of the recent progress has been driven by standard ``benchmarks''. A benchmark is a collection of datasets along with tasks to be learned to perform on the data, specifically designed to measure the performance of ML algorithms in a standardized way. For example, standardized datasets such as MNIST~\cite{deng2012mnist}, CIFAR-10/100~\cite{krizhevsky2009learning}, ImageNet~\cite{deng2009imagenet}, and more recently, VISSL~\cite{goyal2021vissl} have been used to measure the performance of image classification algorithms in computer vision for decades. Other examples include GLUE~\cite{wang2018glue} and SuperGLUE~\cite{wang2019superglue} for natural language processing, and SUPERB~\cite{yang2021superb} and HEAR~\cite{turian2022hear} for human speech/audio processing. Benchmarks are often accompanied by competitions and leaderboards that show a ranking of different ML models based on their performance on those benchmarks. These standardized benchmarks and leaderboards allow for an objective and quantitative comparison of different approaches, the identification of strengths and weaknesses of different methodologies~\cite{olson2017pmlb}, and additionally for an assessment of the overall progress made in the field of research.  Moreover, benchmarks played an important role in the developments and evaluation of some of the recent progress in ML, including the BERT~\cite{devlin2019bert} and GPT-3~\cite{brown2020language} models.

In this paper, we propose BEANS (the BEnchmark of ANimal Sounds), a collection of publicly available bioacoustics datasets along with tasks to be performed on those, specifically designed to measure the performance of ML algorithms in the bioacoustics domain in a standardized manner. The benchmark includes two common bioacoustics tasks, classification and detection, and consists of twelve datasets covering diverse species, including birds, land and marine mammals, anurans, and insects. We run various non-DL and DL algorithms as the baseline on BEANS and show that there is considerable room for improvement, especially for the detection task. We release the entire code and the baseline implementations as open source~\footnote{\url{https://github.com/earthspecies/beans}} to encourage the further development of generic bioacoustic methods.

\section{Benchmark Design}

\begin{figure}[!t]
\begin{center}
\includegraphics[scale=0.38]{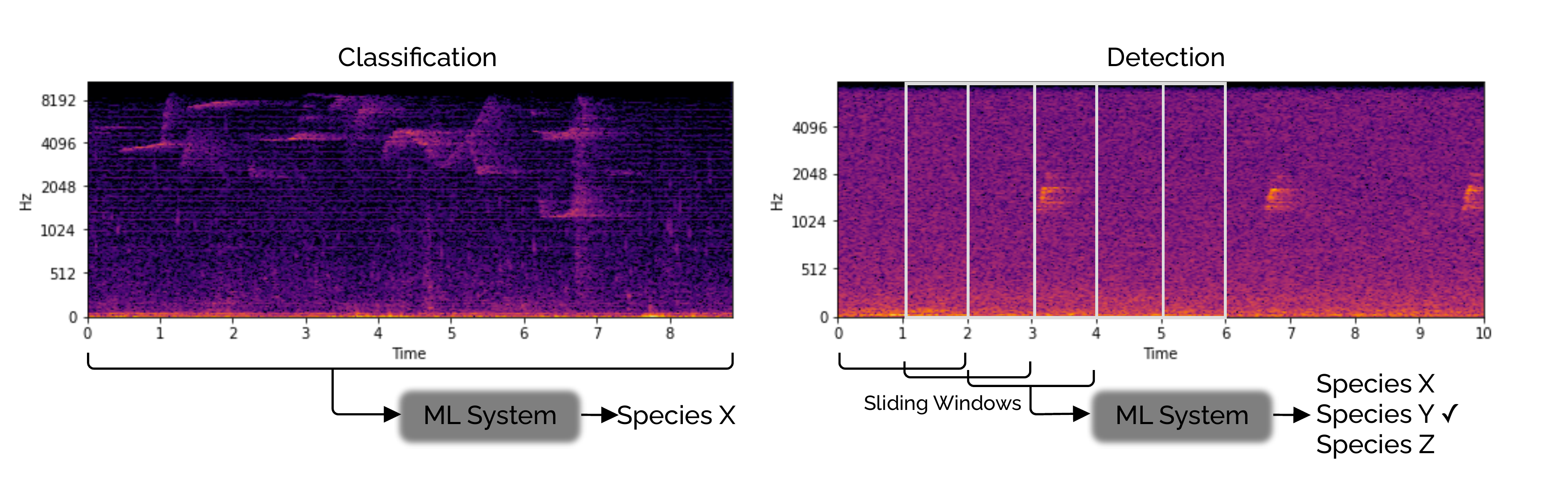} 
\caption{Tasks included in BEANS}
\label{fig:tasks}
\end{center}
\end{figure}

The goal of BEANS is to accurately measure and compare ML models through a collection of bioacoustic datasets covering diverse species. We wish to encourage the development of ML models that work well not only on specific species and dataset, but also on a diverse set of species with as little species-specific modification or training data as possible. Such models are of interest to the bioacoustic community due to better generalizability and lower development cost. 

In this benchmark, we focus on classification and detection tasks, which are the two tasks most commonly considered in the bioacoustics literature~\cite{stowell2022computational}. Classification is a task in which each sample is assigned one or more labels from a set of predefined classes, such as a set of individuals or a set of species. Here, we use a regular single-label, multi-class setting (Figure~\ref{fig:tasks} left).

Detection is a task in which one uses ML algorithms to identify subsections of interest and their properties from long recordings (often obtained from passive acoustic monitoring). We adopt a sliding window approach, as commonly done for detection tasks~\cite{lebien2020pipeline,dufourq2021automated}, where long recordings are broken up into short (potentially overlapping) segments and the ML algorithm makes a prediction per segment (Figure~\ref{fig:tasks} right). In order to address multiple overlapping vocalizations, more than one label can be assigned to a segment, making it a multi-label, multi-class classification setting. 

Due to the design of these tasks, from the perspective of the ML algorithm, the two tasks can be both solved by the same classification model. This homogeneity and simplicity in terms of the structure of the tasks allows the benchmark users to use almost the same algorithm with little modification and encourages the use of generic models.

As metrics for evaluating task performance, we use accuracy $A$ for classification tasks and mean average precision ${\rm mAP}$~\cite{everingham2012pascal} for detection tasks defined below. Specifically, let $N$ be the total number of samples in the dataset, $C$ be the number of classes, and ${\it tp}_c, {\it fp}_c, P_c(r)$ be the number of true positives, false positives, the interpolated precision at recall $r$ for class $c$, respectively. The metrics are then defined as
\begin{eqnarray}
A = \frac{\sum_c ({\it tp}_c + {\it tn}_c)}{CN}, \quad {\rm mAP} = \frac{1}{11 C}\sum_{r \in \{0, 0.1, ..., 1.0\}} P_c(r)
\end{eqnarray}
The benchmark comes with predefined train, validation, and test splits, as well as a baseline implementation, to encourage consistent comparison and reproducibility. Note that we do not encourage the reporting of a single aggregated score across different datasets and tasks for the benchmark, as a single score does not necessarily reflect the difficulties of different tasks, a fact often criticized in the literature~\cite{dehghani2021benchmark}.
 
\section{Datasets}

\begin{table}[!t]
\begin{center}
{\small
\begin{tabular}{@{}llrlrl@{}}
\toprule
Dataset         & Description             & \# Train / Valid / Test$^a$ & \# Labels (type) & Sample Rate & License  \\ \midrule
\multicolumn{5}{l}{Classification}                                                                \\ \midrule
watkins         & aquat. mamm.            & 1017 / 339 / 339        & 31 (species)     &  44.1kHz & free$^b$  \\
bats            & bats                    & 6000 / 2000 / 2000      & 10 (individual)  &  250kHz  & CC-BY-ND \\
cbi             & birds                   &  14207 / 3548 / 3620    &  264 (species)   &  44.1kHz &  CC-BY-NC-SA        \\
dogs            & dogs                    & 415 / 139 / 139         & 10 (individual)  &  44.1kHz & allowed$^c$ \\
humbugdb        & mosquito                & 9293 / 1859 / 1859      & 14 (species)     &  44.1kHz & CC-BY    \\ \midrule
\multicolumn{5}{l}{Detection}                                                                     \\ \midrule
dcase           & birds \& mamm.          & 701 / 233 / 232         & 20 (species)     &  various & CC-BY    \\
enabirds        & birds                   & 230 / 76 / 76           & 34 (species)     &  32kHz   & CC0      \\
hiceas          & cetaceans               & 406 / 134 / 134         & 1 (species)      &  500kHz    & free$^d$    \\
rfcx            & birds \& frogs          & 2835 / 944 / 945        & 24 (species)     &  48kHz   & free$^e$    \\
gibbons         & gibbons                 & 959 / 479 / 639         & 3 (call type)    &  9.6kHz  & CC-NC-SA \\ \midrule
\multicolumn{5}{l}{Auxiliary}                                                                     \\ \midrule
esc-50          & misc. sound            & 1200 / 400 / 400        & 50 (sound type)  &  16kHz    & CC-BY-NC \\
sc              & human                  & 84843 / 9981 / 11006    & 35 (word)        &  44.1kHz  & CC-BY    \\ \bottomrule
\end{tabular}
}
\end{center}
\caption{Datasets included in the benchmark. $^a$ The numbers of samples for classification and the number of 1-minute ``chunks'' for detection (see Section~\ref{sec:experiments} for more details). $^b$ free for personal or academic use, $^c$ academic use allowed through personal correspondence, $^d$ data are free for use without restriction, $^e$ usage allowed for academic research.}
\label{table:datasets}
\end{table}

In this section, we describe the datasets that we included in the benchmark. To make this choice, we surveyed many bioacoustics datasets from the literature and chose 5 datasets for classification and another 5 for detection tasks (Table~\ref{table:datasets}) based on the following criteria:

\begin{itemize}
    \item Availability: Is the dataset publicly and freely available for research purposes? 
    \item Difficulty: Is the dataset moderately difficult for ML algorithms to solve?
    \item Size: is the dataset large enough for ML algorithms to learn meaningful patterns from it? Is the dataset small enough so that the training is within the reach of an average compute budget of typical users (biologists and machine learning researchers)?
    \item Diversity: Does the benchmark represent a diverse collection of sound-making animal species?
\end{itemize}

We also included two ``auxiliary'' datasets that are commonly used to evaluate environmental sound detection and speech classification systems, two domains closely related to bioacoustics, in order to encourage the development of ML models that generalize beyond bioacoustics. These two datasets are not officially part of the benchmark, but benchmark users can choose to report performance numbers on them as a reference.

Below, we describe the specifics of the five classification datasets:

\begin{itemize}
\item {\tt watkins}~\cite{sayigh2016watkins}: The Watkins Marine Mammal Sound Database is a database of marine mammal sounds. We used the preprocessed dataset hosted on the Internet Archive~\footnote{\url{https://archive.org/details/watkins_202104}} which contains the recordings of 32 species from the `Best of cuts' section (except for weddell seal recordings, which had only a few samples). We randomly split the dataset into 6:2:2 train:valid:test portions with stratification. All recordings are resampled to 44.1kHz.
\item {\tt bats}~\cite{prat2017annotated}: The original dataset contains annotated recordings of Egyptian fruit bats ({\it Rousettus aegyptiacus}) vocalizations recorded at a sampling rate of 250kHz. We used the preprocessed dataset consisting of individual calls of up to a few seconds long. The target label is the emitter ID (individual).
\item {\tt cbi}~\cite{cornell2020}: This is the dataset from the Cornell Bird Identification competition hosted on Kaggle. The training set consists of bird recordings uploaded to xeno-canto\footnote{\url{https://xeno-canto.org/}} by volunteer users. Since the test set labels are hidden, we split the train set into 6:2:2 train:valid:test portions in such a way that there is no overlap in recordists between splits.
\item {\tt dogs}~\cite{yin2004barking}: This dataset consists of barks recorded from 10 individual domestic dogs in different situations (disturbance, isolation, and play) originally at 48 kHz and resampled to 44.1kHz. Each recording is annotated with the individual and the situation, but we used the individual as the target label. We randomly split the dataset into 6:2:2 train:valid:test portions with stratification.
\item {\tt humbugdb}~\cite{kiskin2021humbugdb}: HumBugDB is a collection of wild and cultured mosquito wingbeat sounds recorded in various settings (including sounds when the animals were located in cups and under bednets). The purpose is to detect and classify species that can be vectors of diseases such as malaria. We took their species classification dataset, while collapsing any species with 100 or fewer samples into an ``OTHER'' category, resulting in a grouping into 14 classes. We randomly split the dataset into 6:2:2 train:valid:test portions with stratification.
\end{itemize}

The following are the details of the five detection datasets:

\begin{itemize}
\item {\tt dcase}~\cite{morfi2021fewshot}: This is the dataset used for DCASE 2021 Task 5: Few-shot Bioacoustic Event Detection. It contains mammal and bird multi-species recordings annotated with species, onset, and offset times. We repurposed their few-shot development dataset as a (regular) detection dataset by partitioning long recordings into 1-minute chunks, and used the first 60\% for training, the next 20\% for validation, and the final 20\% for testing. We only retained positive (POS) labels. The dataset contains files recorded at various sample rates, but we up/down sampled them all at 16kHz.
\item {\tt enabirds}~\cite{chronister2021annotated}: The dataset contains recordings of bird dawn chorus, annotated with the onset/offset time, the frequency range, and the species. We used the 33 most frequent species labels and treated all infrequent labels as ``OTHERS.'' We partitioned the data set into training, validation, and testing portions as described in {\tt dcase}.
\item {\tt hiceas}~\cite{noaa2022hawaiian}: The dataset consists of a subset of passive acoustic data collected using a multi-channel towed hydrophone during the Hawaiian Islands Cetacean and Ecosystem Assessment Survey (HICEAS) in 2017. We used the human-audible Minke whale ``boing'' vocalization annotations of the data set, rendering this a single-class (binary) detection task. We sampled 1/20th of all the files in the dataset, downsampled them from 500kHz to 22.5kHz, and assigned them to train, valid and test splits randomly with a 6:2:2 ratio. We redistribute the preprocessed dataset as part of the BEANS repository.
\item {\tt rfcx}~\cite{lebien2020pipeline}: This is a dataset of continuous soundscape recordings of 24 species of frogs and birds collected by Rainforest Connection (RFCx). The data were annotated with the onset / offset time, as well as the frequency range. We randomly assigned the files into train, valid, and test splits with a 6:2:2 ratio. We did not use the false positive annotation. 
\item {\tt hainan-gibbons}~\cite{dufourq2021automated}: The dataset contains continuous recordings of Hainan gibbon calls. The data were annotated with onset/offset times and call types (one pulse, multiple pulse, duet). There are a total of 14 files, each corresponding to ~8 hours of recording for a particular day. Due to its large size, we sub-sampled 1/3 of the dataset after splitting into chunks. We used the first 6 files for training, the next 3 files for validation, and the remaining ones for testing.
\end{itemize}

Finally, we included two auxiliary datasets in the benchmark as described above. Note that both are classification tasks.

\begin{itemize}
\item {\tt esc50}~\cite{piczak2015dataset}: A dataset of environmental audio recordings including animal sounds (e.g., dogs, roosters), nature sounds (e.g., rain, sea waves), human (non-speech) sounds (e.g., crying, sneezing), interior/domestic sounds (e.g., door knocks, mouse clicks), and exterior/urban sounds (e.g., helicopter, chainsaw). The dataset is commonly used as a benchmark for environmental sound classification and comes with predefined splits. We used split 4 for validation, split 5 for testing, and the rest for training.
\item {\tt speech-commands}~\cite{warden2018speech}: A dataset of single-word utterances covering 35 English words including digits (e.g., ``zero'', ``one'', ...), commands (e.g., ``yes'', ``no'', ``up'', ``left'', ``stop''), and others (e.g., ``bird'', ``happy'', ``wow''), spoken by multiple speakers. The dataset is commonly used to benchmark speech systems. The dataset comes with predefined train, validation, and test splits. We used version 0.02 of the data set.
\end{itemize}

\section{Experiments}
\label{sec:experiments}

\subsection{Experimental Setup}

We ran a wide range of traditional non-DL and DL algorithms to establish a baseline for BEANS. All the waveforms were mixed to a single channel (mono) with 16 bit depth. For classification, each waveform was padded with silence at the end if it was shorter than the minimum duration threshold set for the dataset, or truncated if it was longer. For detection, a sliding window algorithm was applied to partition the chunks into instances. The length of sliding windows was 2 seconds for {\tt dcase} and {\tt enabirds}, 10 seconds for {\tt hiceas} and {\tt rfcx}, and 4 seconds for {\tt hainan-gibbons}. An instance is marked positive if the amount of overlap with any annotation is more than 20\%. 

We consider the following non-DL algorithms, often used for classification of tabular data:

\begin{itemize}
    \item Logistic regression (LR)
    \item Support vector machine (SVM)
    \item Decision tree (DT)
    \item Gradient-boosted decision tree (GBDT)
    \item XGBoost (XGB, \cite{chen2016xgboost})
\end{itemize}

We used the official library for XGBoost~\footnote{\url{https://xgboost.readthedocs.io/en/stable/}} and the implementations in \texttt{scikit-learn}~\cite{pedregosa2011scikit} for the other four algorithms. We first obtained 20 MFCCs features from power mel-spectrogram~\cite{yang2022torchaudio}. The length of the FFT window and the hop length were chosen as 50ms and 10ms, respectively. We computed four summary statistics, mean, standard deviation, min, and max of each MFCC dimension over time, resulting in a 80-dimension feature vector per sample.

Furthermore, we consider the following DL-based models. They are all based on convolutional neural networks (CNNs), a DL architecture which has been successfully applied in most fields of ML, and also the fields of bioacoustics~\cite{ruff2020automated}, environmental sound classification~\cite{boddapati2017classifying}, and human speech~\cite{kriman2020quartznet}.

\begin{itemize}
    \item ResNet~\cite{he2016deep} (ResNet18, ResNet50, ResNet152, both random and pretrained weights on ImageNet)
    \item VGGish~\cite{hershey2017cnn}, a VGG-like architecture~\cite{simonyan2015very} pretrained on a large YouTube audio dataset
\end{itemize}

The VGGish model has been widely used as a strong baseline model for various audio classification and detection tasks~\cite{stowell2022computational}. The input to those models were power mel-spectrograms computed with \texttt{torchaudio}~\cite{yang2022torchaudio}, using the same parameters as above. We downsampled the input to 16kHz for VGGish due to its preprocessing pipeline. We used the ResNet models implemented in \texttt{torchvision}~\cite{marcel2010torchvision}. The representations were average-pooled before being fed to the classification layer. For classification tasks, a linear and a softmax layer were added on top of the classification layer (for ResNet) and the embedding layer (for VGGish), and the entire network was optimized using a cross-entropy loss function. For detection tasks, the softmax layer was replaced with a sigmoid layer and the network was optimized with a binary cross-entropy loss function.

Each model was fine-tuned on the training portion of each dataset using the Adam \cite{kingma2014adam} optimizer with $\beta_1 = 0.9, \beta_2 = 0.999, \epsilon = 1.0\times10^{-8}$. We swept the learning rate over $1.0\times10^{-5}, 5.0\times10^{-5}, 1.0\times10^{-4}$ for 50 epochs, and picked the best model based on the validation metric.

\subsection{Main Results}

\begin{table}[!t]
\begin{center}
{\small
\begin{tabular}{@{}lllllllllllll@{}}
\toprule
       & \multicolumn{5}{l}{Classification}    & \multicolumn{5}{l}{Detection}          & \multicolumn{2}{l}{Auxiliary} \\ \midrule
       & wtkn  & bat   & cbi   & hbdb  & dogs  & dcase & enab  & hiceas & rfcx  & gib   & esc           & sc            \\ \midrule
lr       & 0.776          & 0.661          & 0.156          & 0.751          & 0.885          & 0.143          & 0.247          & 0.221          & 0.030          & 0.006          & 0.428          & 0.535           \\
svm      & \textbf{0.870} & {\ul 0.720}    & 0.139          & {\ul 0.779}    & \textbf{0.914} & 0.146          & 0.299          & 0.218          & 0.038          & 0.039          & 0.478          & 0.572           \\
dt       & 0.661          & 0.474          & 0.025          & 0.700          & 0.626          & 0.095          & 0.183          & 0.223          & 0.009          & 0.007          & 0.240          & 0.230           \\
gbdt     & 0.758          & 0.674          & 0.038          & 0.759          & 0.827          & 0.104          & 0.235          & 0.216          & 0.009          & 0.007          & 0.338          & 0.481           \\
xgb      & 0.808          & 0.692          & 0.097          & 0.772          & 0.842          & 0.124          & 0.270          & 0.214          & 0.012          & 0.007          & 0.403          & 0.525           \\ \midrule
rn18     & 0.752          & 0.443          & 0.357          & 0.697          & 0.662          & 0.161          & 0.325          & 0.280          & 0.064          & 0.164          & 0.500          & 0.926           \\
rn50     & 0.799          & 0.548          & 0.295          & 0.696          & 0.633          & 0.183          & 0.282          & {\ul 0.304}    & 0.055          & 0.215          & 0.235          & 0.936           \\
rn152    & 0.743          & 0.483          & 0.330          & 0.645          & 0.511          & 0.154          & 0.280          & 0.255          & 0.069          & {\ul 0.248}    & 0.365          & 0.929           \\
rn18p    & 0.735          & 0.532          & 0.509          & 0.649          & 0.705          & 0.223          & {\ul 0.462}    & 0.262          & 0.079          & \textbf{0.316} & {\ul 0.590}    & 0.936           \\
rn50p    & 0.735          & 0.560          & {\ul 0.548}    & 0.673          & 0.763          & 0.178          & 0.424          & 0.284          & {\ul 0.087}    & 0.155          & 0.545          & {\ul 0.946}     \\
rn152p   & 0.720          & 0.544          & \textbf{0.573} & 0.662          & 0.741          & 0.198          & 0.429          & 0.273          & 0.085          & 0.230          & 0.540          & {\ul 0.946}     \\
vggish   & {\ul 0.847}    & \textbf{0.743} & 0.440          & \textbf{0.808} & {\ul 0.906}    & \textbf{0.335} & \textbf{0.535} & \textbf{0.463} & \textbf{0.140} & 0.150          & \textbf{0.705} & \textbf{0.948} \\ \bottomrule
\end{tabular}
}
\end{center}
\caption{Main results. As measure for task performance, we used accuracy for classification and auxiliary tasks, and mean average precision (mAP) for detection tasks. The best and second best performing models are highlighted and underlined, respectively, for each dataset.}
\label{table:results}

\end{table}

The main results are shown in Table~\ref{table:results}. In terms of model performance, VGGish works best overall, followed by pretrained ResNet models. Among the non-DL based ML models, SVM shows surprisingly good performance and beats neural models in some classification tasks. Note that ResNets are pretrained on ImageNet, while VGGish is pre-trained on a YouTube dataset~\cite{hershey2017cnn}. We hypothesize that if we were to pretrain a ResNet on a large-scale audio datasets such as AudioSet~\cite{gemmeke2017audioset}, that the performance attainable by ResNets could be better. We leave this for future work.

Cornell Bird Identification (CBI) is the most challenging classification task in BEANS, due to the way the data is split (with no overlapping recordists) and the sheer number of target labels (264 species), which is close to how machine learning models are used in some real-world bioacoustic settings.

In terms of tasks, performance on some detection datasets (e.g., {\tt rfcx}, {\tt gibbons}) was lower than others. This might be due to the fact that a typical detection dataset is very sparse, meaning that only a minor portion of a typical recording is animal vocalization and has fewer training annotations per class than classification. Modern regularization and data augmentation techniques, such as mixup~\cite{zhang2018mixup} and SpecAugment~\cite{park2019specaugment} may help improve the results.

For most of the datasets contained in BEANS, the performance of our baselines is well below 90\%, which means that there is large room for improvement for future, more specialized ML algorithms. 

Note that the way we padded the instances could inadvertently affect the classification performance if there exists systematic difference in instance lengths between classes irrespective of true vocalization lengths, although we didn't find any evidence that padding is leading to an overestimation of classification performance.

\section{Discussion}

\paragraph{Sample rate} Bioacoustics data can have a wide range of sample rates, which can be challenging for ML models that are usually trained on data captured at a single and fixed sample rate. Even within BEANS, vocalizations range from human-audible calls that can be captured with a sampling rate around 10kHz to ultrasonic echolocation of bats that require sampling rates of 250kHz. Methods that can be adopted to different sampling frequencies (e.g., S4~\cite{gu2022efficiently}) could be a promising future direction.

\paragraph{Overlapping calls} Currently, we cannot address the cases where multiple individuals are vocalizing at the same time unless those individuals are explicitly and separately annotated. Sound separation models (e.g., BioCPPNet~\cite{bermant2021biocppnet}) are a promising preprocessing step to address this issue.

\paragraph{Bias} Some animal species and/or geographical locations can be over- or under-represented in a dataset. For example, North American birds are over-represented in BEANS, included in at least three datasets ({\tt cbi}, {\tt dcase}, and {\tt enabirds}). We acknowledge that the curating process of datasets is inherently subjective~\cite{shankar2017no, raji2021ai} and wish to address this bias in future iterations of this benchmark.

\paragraph{Limitations} We acknowledge that a benchmark is only a proxy for progress that we measure and can face construct validity issues~\cite{raji2021ai,bowman2021what}. We also note the inherent fragility of the ML benchmarking process~\cite{dehghani2021benchmark}. Therefore, we in particular do not recommend reporting a single metric averaged over datasets and tasks as a performance measure of an ML algorithm on this benchmark. Instead, BEANS is intended as a diagnostic tool to encourage the development of generic, well-balanced bioacoustics approaches.

\bibliography{neurips_2022}
\bibliographystyle{plain}

\end{document}